# Apparatus for measuring the uniformity of the optical transmittance of a semispherical surface at normal incidence

Shenghao Wang, Shijie Liu\*, Jianda Shao\*, Tianzhu Xu, Qi Lu, Shen Qi, Minghui Feng, Long Zhang

*Key Laboratory of Materials for High Power Laser, Shanghai Institute of Optics and Fine Mechanics, Chinese Academy of Sciences, Shanghai, 201800, China.*
*\*Corresponding authors: shijieliu@siom.ac.cn; jdShao@siom.ac.cn.*

**In this manuscript, we build an apparatus for measuring the optical transmittance and its uniformity for a semi-spherical surface at normal incidence; the system is primarily comprised of a traditional double-beam photometric framework and a novel custom-made mechanical structure with multidimensional degrees of freedom. During the measurement process, a key aligning step is adopted to guarantee that the center point of the semi-spherical surface stands still in the light beam while scanning the hemispherical optical element point by point around the horizontal and vertical axes, which ensures that the laser beam is always normally incident onto the surface of the hemisphere. The experimental results show that the uniformity of the optical transmittance for a semi-spherical optical glass can be successfully characterized by the system, with a 3-cycle repeatability error of 0.026% being demonstrated. Our system solves the problem of traditional spectrophotometers when measuring the spectral property of a hemispherical surface and thus can be popularized in similar applications.**

http://dx.doi.org/10.1364/AO.99.099999

## 1. INTRODUCTION

Optical glasses with a semi-spherical surface have been widely used in many optical systems and their spectral properties, such as transmittance, reflectance and uniformity, in the whole aperture at normal incidence is a highly important indicator when evaluating the system performance [1-7]. Currently, the mostly popular techniques for measuring the transmittance and reflectance of optical glass involve spectrophotometry [8-10], methods based on the use of a CCD-array spectrometer [11, 12] and a Fourier transform spectrometer [13, 14]. However, to obtain an accurate measurement result when performing tests using the aforementioned techniques, typically, the sample should be well-prepared with a certain external size (normally 50×50 mm) and shape (the surface under inspection is usually a plane) [15-17]; Consequently, in the case of normal incidence, the transmittance and its uniformity over the whole aperture of a large optical element with a semi-spherical surface cannot be precisely characterized using commercially available machines (the optical transmittance changes when the beam is not normally incident on the surface of the sample).

With the aim of measuring the transmittance and its uniformity of a large optical glass with a semi-spherical surface at normal incidence, in this manuscript, we build a dedicated apparatus for measuring the spectral characteristic of a large semi-spherical surface.

## 2. MEASUREMENT PRINCIPLE

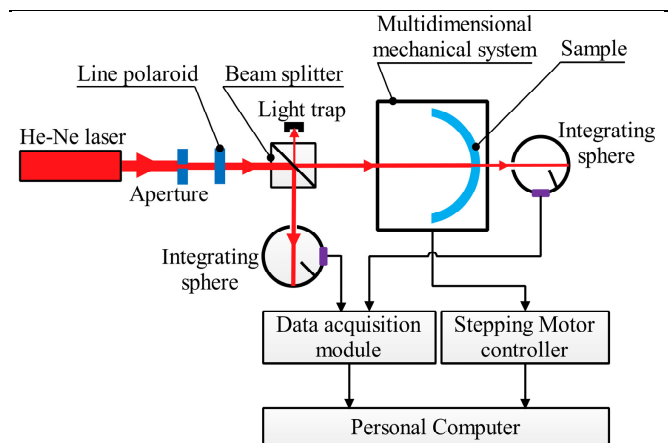

**Fig. 1.** Framework of the system used for measuring the uniformity of the optical transmittance for a semi-spherical surface at normal incidence.

The measurement is achieved herein based on use of the traditional double-beam photometric framework and a novel custom-made mechanical structure with multidimensional degrees of freedom. The system is shown in Fig. 1; it consists of a laser source, an unpolarized beam splitter, which generates the double light beams (in order to eliminate the stray light, a light trap is used to block the unnecessary light beam), a customized multidimensional mechanical system and two integrating spheres (including photoelectric detectors). The adoption of a double-beam can successfully eliminate the negative effect of laser power fluctuation on measurement accuracy; a detailed measurement theory for the optical transmittance of a plane sample at a single sampling point using the double-beam photometry was described in the literatures [8, 18-20].

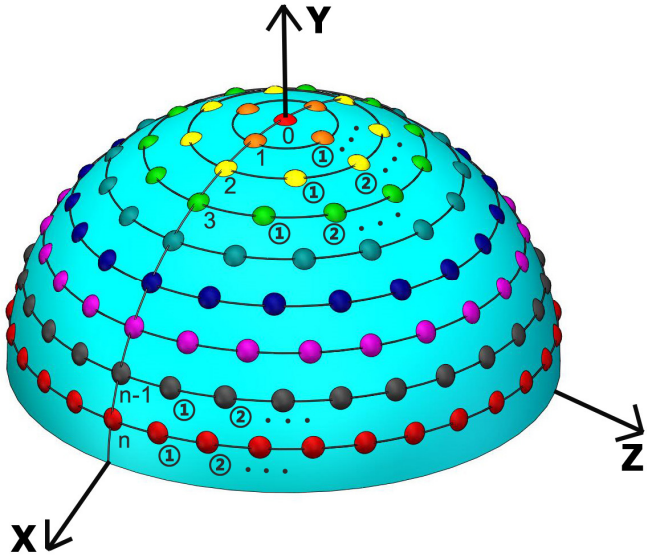

**Fig. 2.** Distribution of the sampling points on the semi-spherical surface.

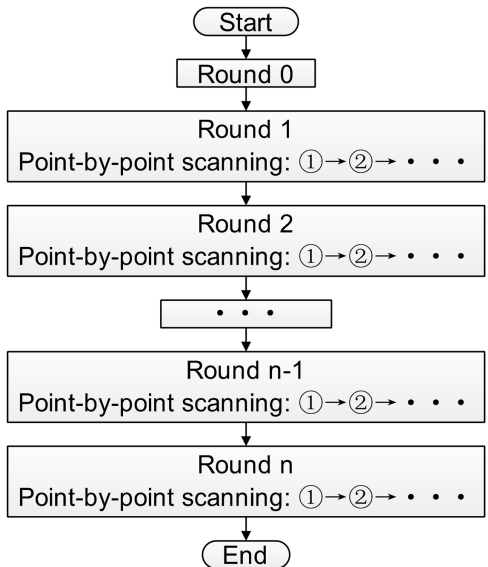

**Fig. 3.** Flow chat of the scanning process.

The custom-made mechanical structure works in this study to align the semi-spherical sample in such a way that the laser beam is always normally incident onto the surface of the hemisphere, while the sample is rotated for scanning the semi-spherical sample point by point all over the surface. In the process of alignment, the center point of the semi-spherical surface is moved to be located in the rotation axes of the motorized stages, which scan the hemispherical sample around the Z- & Y-axes, with the light beam reflected by the inner surface of the semi-spherical sample used to judge whether the laser beam is normally incident onto the surface of the hemisphere (this can be done by mounting an aperture in the incident beam path; if the laser beam is normally incident onto the surface of the hemisphere, the reflected light beam will also pass through the aperture).

Fig. 3 is the flow chat of the scanning process, we described in detail as follows: during the scanning, as shown in Fig. 2, first the semi-spherical optical glass is moved such that the laser beam is incident onto the initial sampling point (vertex of the semi-spherical surface, labeled as 0 in Fig. 2); then, the sample is rotated to a predetermined angular interval around the Z-axis, followed by rotation of the Y-axis step by step over 360°; in this way, all the sampling points located in the 1st round of the surface (see the orange dots) can be measured; by repeating the aforementioned process, measurement of all the sampling points (see dots of the same color) equally spaced on each round (2nd, 3rd ⋯ nth) is realized; finally, the uniformity of the optical transmittance for a semi-spherical surface at normal incidence can be appropriately characterized; it is noted that the number of sampling points in different rounds should be in proportion to the perimeter of the corresponding round.

## 3. EXPERIMENTAL SETUP
### A. Physical Map of the Measuring System

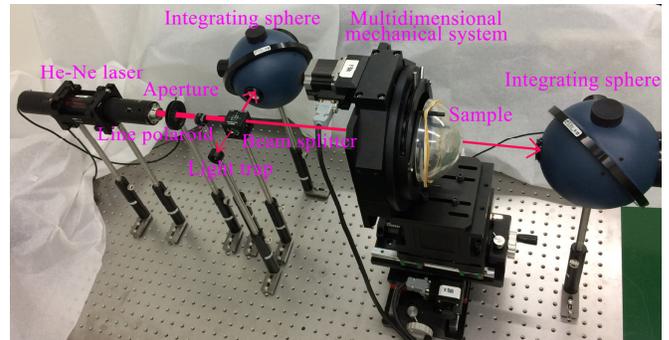

**Fig. 4.** Physical map of the system used for measuring the uniformity of the optical transmittance for a semi-spherical surface.

As shown in Fig. 4 is the physical map of the system for measuring the optical transmittance and its uniformity for a semi-spherical surface at normal incidence, it is primarily comprised of a stabilized He-Ne laser source, an aperture, a line polaroid, an unpolarized beam splitter, a light trap, a customized multidimensional mechanical system, an infrared optical glass with a semi-spherical surface, two integrating spheres (including photoelectric detectors), a data acquisition module, a stepping motor controller and a personal computer. The stabilized He-Ne laser source emits a beam with a power higher than 1.2 mW at a center wavelength of 632.99 nm, with a frequency stabilization of approximately ±2 MHz, and a guaranteed intensity stabilization better than ±0.2%. The line polaroid has an effective wavelength range of 550 - 1500 nm, and an extinction ratio that is higher than 1000:1. The unpolarized beam splitter can be applied from 400 nm to 700 nm with a splitting ratio of approximately 50:50. The integrating sphere has an inner diameter of 5.3 inches, with a silicon detector mounted on the output port; its spectral response wavelength range is 200 - 1100 nm, and the power detecting range is between 300 nW and 1 W. The data acquisition module can be used to simultaneously measure the signals transmitted from the two photoelectric detectors via the dual input channels.

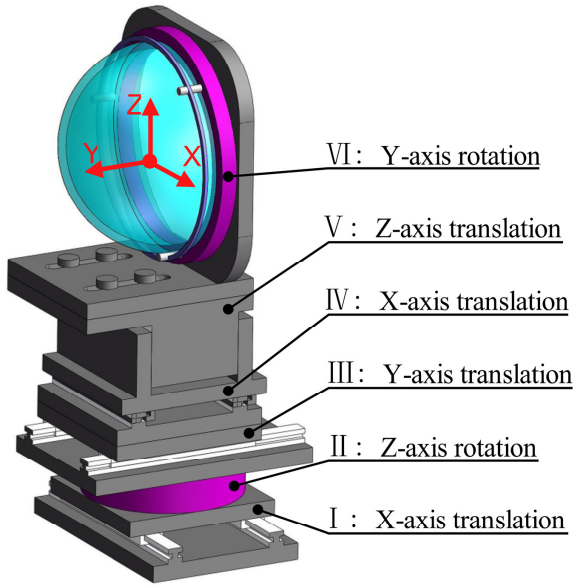

**Fig. 5.** 3D Structure of the multidimensional mechanical system.

The 3D structure of the multidimensional mechanical system is illustrated in Fig. 5; from bottom to top, the constituent elements are, successively, an X-axis translation stage (I), Z-axis motorized rotation stage (II), Y-axis translation stage (III), X-axis translation stage (IV), Z-axis translation stage (V), Y-axis motorized rotation stage (VI) and clamping device; the two motorized stages are operated by the stepping motor controller, and a positioning accuracy of ±0.001° is guaranteed; the sample under inspection is an infrared optical glass with a semi-spherical surface and a diameter of approximately 12 cm.

The monochromatic light beam emitted by the He-Ne laser source first passes through the optical aperture and subsequently penetrates the line polaroid; next, a reference light beam and testing beam are simultaneously generated under the action of the unpolarized beam splitter; the reference integrating sphere (equipped with a photoelectric detector) works here for collection of the reference light beam, while the testing integrating sphere (also including a photoelectric detector) captures the testing beam after it penetrates through the inspected sample; the photo signal detected by the two detectors is first photoelectrically converted, and then digitized by the dual channel data acquisition module. The multidimensional mechanical system is used here to both align the semi-spherical surface and scan the different sampling points. The personal computer is utilized to remotely control the motorized stages and operate the data acquisition module through serial ports; LabVIEW software is developed to implement the desired functionality such as system initialization, sample scanning, data acquisition and post-processing, graphic display and data storage.

## B. Working Principle and Testing Procedure of the Setup

As we mentioned above, the working theory for the double-beam photometry can be found elsewhere [8, 18-20]; therefore, in this study, we emphasize the working principle of the multidimensional mechanical system for aligning the semi-spherical optical glass and scanning over the whole surface.

As shown in Fig. 5, stage I is used to move the rotation axis of stage II such that its rotation axis and the light beam lie in the same plane; stage III works here for adjusting the Y-axis position of the center point of the semi-spherical surface to be located in the rotation axis of stage II; stage IV and stage V are used here to move the position of the center point of the semi-spherical surface into the light beam; stage II and stage VI are utilized to rotate the sample around the Z- and Y-axes, respectively.

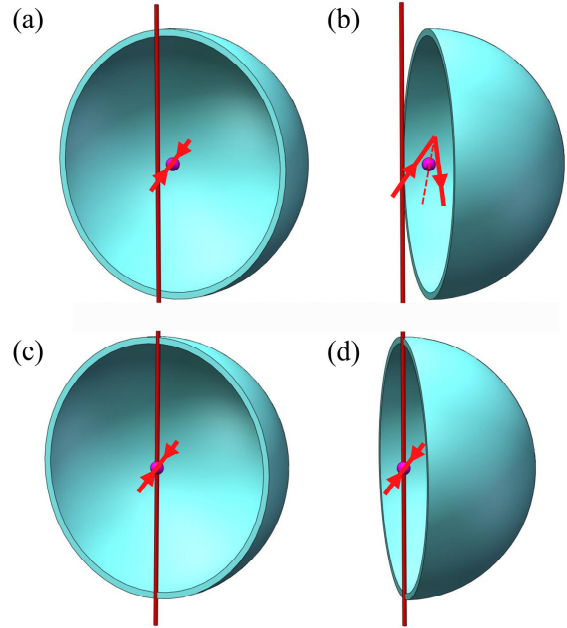

**Fig. 6.** Schematic diagram for the alignment; the violet dot is the center point of the semi-spherical surface, and the horizontal red lines with arrows show the incident and reflected light beams; the vertical line is the rotation axis of stage II. (a, b) When the center point of the semi-spherical surface is not located in the rotation axis of stage II, the reflected light beam and incident beam will not be collinear once the semi-spherical surface is rotated by a certain angle, (c, d) when the center point of the semi-spherical surface is perfectly located in the rotation axis of stage II, the reflected light beam and the incident beam will always be collinear during rotation of the semi-spherical surface.

When the rotation axis of stage II and the light beam lie in the same plane, and the center point of the semi-spherical surface is located in the light beam, as shown in Fig. 6(a) and Fig. 6(b), the light beam reflected by the sample and the incident beam will not be collinear following rotation of stage II by a certain angle because the center point of the semi-spherical surface is not located in the rotation axis of stage II; by moving stage III to the position shown in Fig. 6(c) and Fig. 6(d), where the light beam reflected by the inner surface of the sample and the incident beam are collinear, it can be found that the two light beams will always lie on the same line while scanning the semi-spherical surface by rotating either stage II or stage VI. In this case, the goal that the center point of the semi-spherical surface stands still in the light beam while scanning the hemispherical optical element around the Z- & Y-axes is achieved, which ensures that the laser beam is always normally incident onto the surface of the hemisphere when rotating either stage II or stage VI.

The automatic scanning of the semi-spherical optical glass can be begun after aligning the sample; Fig. 7 demonstrates point by point scanning over the whole surface of a semi-spherical optical glass; first the semi-spherical optical glass is moved to the initial sampling point (vertex of the semi-spherical surface) by rotating stage II and stage VI, and, after data acquisition, the sample is rotated by a predetermined angular interval using stage II; then, via step by step rotation of stage VI across 360°, all the sampling points located in the first round of the surface can be measured; by repeating the aforementioned process, measurement of the sampling points equally spaced across all the selected rounds is completed; therefore, the optical transmittance and its uniformity for a semi-spherical surface at normal incidence can be characterized.

S4: Remove the light trap away from the beam path, and let the integrating sphere directly collect the testing beam without penetrating through the sample; then, simultaneously capture the power of both the reference light beam and testing beam, denoted respectively as $I_{ref}^1$ and $I_{ref}^2$ and compute the intensity ratio $k_0$ as

$$k_0 = \frac{I_{ref}^2 - I_{dark}^2}{I_{ref}^1 - I_{dark}^1}. \quad (1)$$

S5: Mount the sample onto the multidimensional mechanical system with the aid of a clamping device, and rotate stage II to a certain angle; then, move stage III to the position where the reflected beam is collinear with the incident beam; next, move the sample to the initial position $P_1$ using stage II and stage VI.

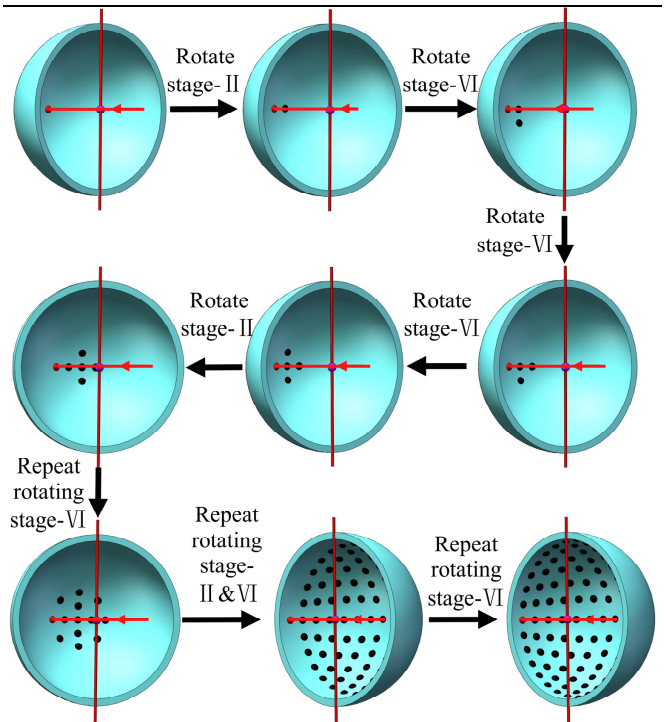

**Fig. 7**. Demonstration of point by point scanning for a semi-spherical optical glass; the horizontal red line with an arrow shows the incident light beam; the vertical line is the rotation axis of stage II; the red and dark dots show the sampling points under measurement and those that have already been measured, respectively.

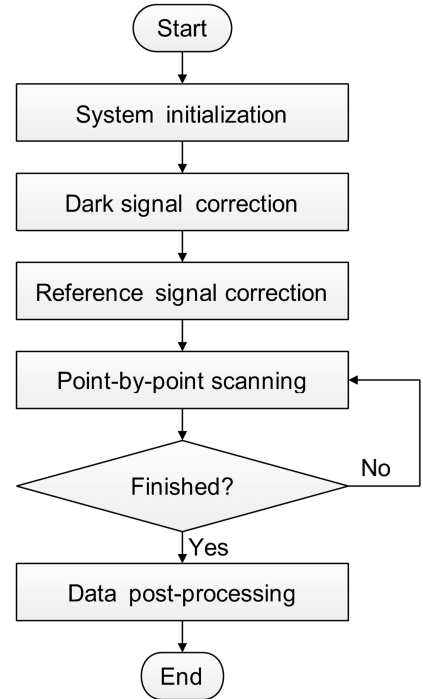

**Fig. 8**. Flow chat of the testing procedure.

Fig. 8 is flow chat of the testing procedure used for obtaining the transmittance and its uniformity for a semi-spherical surface at normal incidence, we described in detail as follows:

S1: Align the laser beam parallel to the plane of the optical platform, and then, remove the other parts of the multidimensional mechanical system above stage II; then, move stage I such that the rotation axis of stage II and the laser beam lie in the same plane.

S2: Assemble the multidimensional mechanical system, making sure that the light beam passes through the rotation axis of stage VI by moving stage IV and stage V.

S3: Place a light trap in the beam path between the line polaroid and unpolarized beam splitter; then, simultaneously capture the power of the reference light beam and the testing beam via the dual channel data acquisition module, denotes respectively as $I_{dark}^1$ and $I_{dark}^2$.

S6: Simultaneously acquire the intensity of the reference and testing light beam, denoted respectively as $I_{sig}^1$ and $I_{sig}^2$, and compute the intensity ratio $k_{P_1}$ as

$$k_{P_1} = \frac{I_{sig}^2 - I_{dark}^2}{I_{sig}^1 - I_{dark}^1}. \quad (2)$$

S7: Scan the semi-spherical surface successively through the sampling points $P_2$, $P_3$ ⋯ $P_{n-1}$ and $P_n$ by rotating stages II and VI in the predetermined manner, as demonstrated in Fig. 7, and repeat step S6 at each sampling point; compute the intensity ratio $k_{P_2}$, $k_{P_3}$ ⋯ $k_{P_{n-1}}$ and $k_{P_n}$ corresponding to each sampling point.

S8: Use equation (3) to calculate respectively the transmittance $T_{P_1}$, $T_{P_2}$, $T_{P_3}$ ··· $T_{P_{n-1}}$ and $T_{P_n}$ of the sample corresponding to the sampling points $P_1$, $P_2$, $P_3$ ··· $P_{n-1}$ and $P_n$:

$$T_{P_i} = \frac{k_{P_i}}{k_0}. \quad (3)$$

where $k_{P_i}$ and $T_{P_i}$ represent the intensity ratio and the optical transmittance at the sampling point of $P_i$, respectively.

S9: Plot the contour of the transmittance over the semi-spherical surface using the data obtained at all the sampling points.

## 4. EXPERIMENTAL RESULTS AND DISCUSSION

Fig. 9(a) shows the distribution of the sampling points on the semi-spherical surface of the infrared optical glass; in total, 181 sampling points were measured in the experiment; beside the initial position (as demonstrated in Fig. 9(a) by the central point), the other sampling points are located across 9 rounds over the semi-spherical surface; in the 1st, 2nd, 3rd ··· 9th round, the selected sampling points are 4, 8, 12 ··· 36, respectively; all these points are equally spaced across the rounds. Fig. 9(b) shows the contour of the normalized optical transmittance measured over the semi-spherical surface; we can see that the infrared semi-spherical optical glass shows considerably good uniformity apart from the defects located on the upper part.

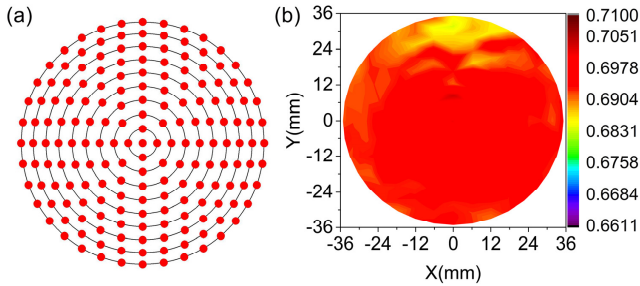

**Fig. 9.** (a) Distribution of the 181 sampling points over the semi-spherical surface, (b) contour of the normalized optical transmittance.

**Table 1.** Numerical statistics for the 181 sampling points

| Number | Category | Value |
|---|---|---|
| 1 | Mean | 0.6938 |
| 2 | Maximum | 0.7032 |
| 3 | Minimum | 0.6838 |
| 4 | Peak-valley | 0.0194 |
| 5 | (Peak-valley) / Mean | 0.0280 |
| 6 | Standard deviation | 0.0033 |
| 7 | Coefficient of variation | 0.0048 |

Tab. 1 shows the numerical statistics for the optical transmittance measured at the 181 selected sampling points across the semi-spherical surface of the infrared optical glass.

It should be noted that the custom-built measuring system was calibrated by a standard sample with known optical transmittance (measured by a commercial spectrophotometer: PerkinElmer Lambda 1050) before the aforementioned measurement.

Fig. 10 shows the data measured for repetitive scans and a control experiment; Fig. 10(a), Fig. 10(b) and Fig. 10(c) show the results obtained from three successive measurements; the repeatability error is written as

$$\zeta = \frac{\sum_{k=1}^{181} \chi_{p_k}}{181} = 0.026\%. \quad (4)$$

where $\chi_{p_k}$ is the repeatability error at the sampling point $p_k$, which is computed as follows:

$$\chi_{p_k} = \frac{\sqrt{\sum_{i=1}^{3}\left(T_i - \sum_{i=1}^{3} T_i / 3\right)^2 / 3}}{\sum_{i=1}^{3} T_i / 3}. \quad (5)$$

In equation (5), $T_i$ represents the optical transmittance of the sample at sampling point $p_k$ in the $i$ th measurement.

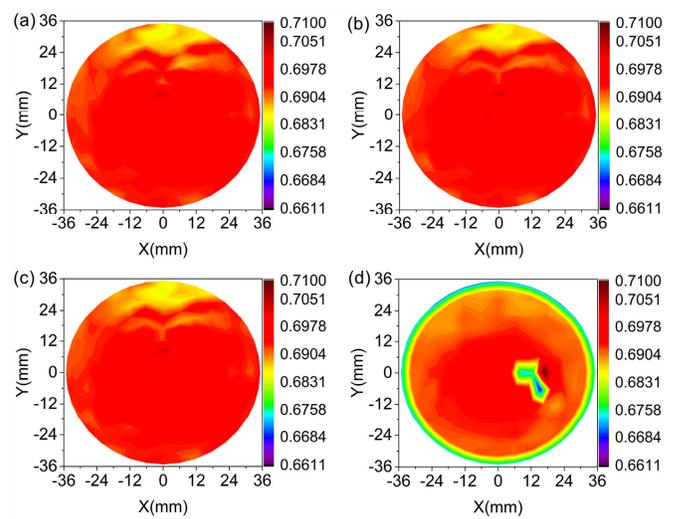

**Fig. 10.** (a-c) Normalized optical transmittance measured for three successive scans, (d) the obtained data when the center point of the semi-spherical surface is not located in the rotation axis of stage II.

Fig. 10(d) shows a control experiment when the position of the center point of the semi-spherical surface is not located in the rotation axis of stage II; this property is achieved by moving stage III to a position where the reflected light beam and the incident beam are not collinear once stage II is rotated by a certain angle, as shown in Fig. 6(b). In contrast to Fig. 10(c), it can be seen clearly from Fig. 10(d) that a non-negligible measuring error appears in the outer ring of the semi-spherical surface (this phenomenon can be explained by the relatively large angular deviation for the incident angle in the outer ring), this experimental control confirms the importance of aligning the semi-spherical surface such that the incident laser beam is always normally incident onto the surface of the hemisphere during scanning.

## 5. CONCLUSIONS

We built an apparatus for measuring the optical transmittance and its uniformity for a semi-spherical surface at normal incidence; the experimental results show that the uniformity of the optical transmittance for a semi-spherical optical glass at normal incidence can be successively characterized by the system, with a 3-cycle repeatability error of 0.026% being demonstrated. Our system solves the problem of traditional techniques when measuring the spectral property of a hemispherical surface and thus can be popularized in similar applications.

Finally, we note that in the current configuration, the optical transmittance can only be measured for a part of the semi-spherical surface due to the incompact design of the mechanical part (the incident beam is blocked when rotating stage II by a large angle). In the future, the mechanical system will be optimized such that most of the area of the sample can be measured, and meanwhile, measurement of the spectral property of an aspheric surface would be considered.

**Funding Information.** National Natural Science Foundation of China (61705246, 11602280).